\documentclass[twocolumn]{aastex63}
\usepackage[all]{hypcap}
 
\usepackage{amsmath}
\DeclareRobustCommand{\okina}{%
  \raisebox{\dimexpr\fontcharht\font`A-\height}{%
    \scalebox{0.8}{`}%
  }%
}

\graphicspath{{./}{figures/}}

\begin{document}

\title{Spatial and Temporal Distribution of Nanoflare Heating During Active Region Evolution}

\correspondingauthor{Biswajit Mondal}
\email{biswajit70mondal94@gmail.com, biswajit.mondal@nasa.gov}
\author[0000-0002-7020-2826]{Biswajit Mondal}
\affiliation{NASA Postdoctoral Program, NASA Marshall Space Flight Center, ST13, Huntsville, AL, USA}
\author[0000-0003-2255-0305]{James A Klimchuk}
\affiliation{NASA Goddard Space Flight Center, Heliophysics Science Division, Greenbelt, MD 20771, USA}
\author[0000-0002-5608-531X]{Amy R. Winebarger}
\affiliation{NASA Marshall Space Flight Center, ST13, Huntsville, AL, USA}
\author[0000-0002-4454-147X]{P. S. Athiray}
\affiliation{Center for Space Plasma and Aeronomic Research, The University of Alabama in Huntsville, Huntsville, AL, USA}
\affiliation{NASA Marshall Space Flight Center, ST13, Huntsville, AL, USA}
\author[0000-0002-7290-0863]{Jiayi Liu}
\affiliation{Institute for Astronomy, University of Hawai$\okina$i at M\={a}noa, 2680 Woodlawn Dr., Honolulu, HI 96822, USA}

\begin{abstract}

Nanoflares are believed to be key contributors to heating solar non-flaring active regions, though their individual detection remains challenging.
This study uses a data-driven field-aligned hydrodynamic model to examine nanoflare properties throughout the lifecycle of AR12758. 
We simulate coronal loop emissions, where each loop is heated by random nanoflares depending on the loop parameters derived from photospheric magnetograms observed by SDO/HMI. 
Simulated X-ray flux and temperature can reproduce the temporal variations observed by Chandrayaan-2/XSM. Our findings show that high-frequency nanoflares contribute to cool emissions across the AR, while low- and intermediate- frequency primarily contribute to hot emissions. During the emerging phase, energy deposition is dominated by low-frequency events. Post-emergence, energy is deposited by both low- and intermediate-frequency nanoflares, while as the AR ages, the contribution from intermediate- and high-frequency nanoflares increases.
The spatial distribution of heating frequencies across the AR reveals a clear pattern: the core of the active region spends most of its time in a low-frequency heating state, the periphery is dominated by high-frequency heating, and the region between the core and periphery experiences intermediate-frequency heating.

\end{abstract}

\section{Introduction}

Exploring the mechanism(s) responsible for heating in non-flaring active regions continues to be a dynamic and engaging area of study in heliophysics. 
It is widely accepted that magnetic fields are the primary drivers of coronal heating. The movement of magnetic field foot-points, caused by the photospheric convective motion, can lead to either the generation of waves or the quasi-static buildup of magnetic energy, depending on the timescale of the motion~\citep{Klimchuck_2006SoPh}. 
Heating through the dissipation of magnetic energy (e.g.,~\citealp{parker_1988}) or dissipation of waves (e.g.,~\citealp{Alfven_1947}) can give rise to impulsive heating events, known as nanoflares~\citep{klimchuk_2015RSPTA}. 
The frequency and magnitude of nanoflares are key factors in determining their contribution to the coronal heating budget. 
Nanoflares are classified into three categories based on their heating frequency: High-Frequency (HF), Low-Frequency (LF), and Intermediate-Frequency (IF) nanoflares.

In the case of HF nanoflares, the time between successive events is much shorter than the plasma cooling timescales. Conversely, for LF nanoflares, the time between events is significantly longer than the plasma cooling time. IF nanoflares fall between these two extremes, with the time between events being comparable to the plasma cooling time.

{The corona is comprised of an enormous number of quasi-independent magnetic strands. There of order 100,000 of them in a typical active region~\citep{klimchuk_2015RSPTA}. 
Nanoflares heat both the diffuse component of the corona, which accounts for the majority of the total emission (specifically in EUV wavelength), and the observationally distinct loops that represent localized ``storms" of nanoflares~\citep{Klimchuk_2023FrP}.
A central problem in coronal physics is understanding the frequency of nanoflare occurrences within individual strands.}
In the absence of direct detection of the individual nanoflares, we often use their observable collective properties in EUV and X-ray emissions; e.g., existence of very hot ($\sim$10 MK) plasma~\citep{Brosius_2014ApJ, Ishikawa_2017NatAs}, measuring differential emission measure (DEM) distribution~\citep{Reale_2009ApJ, Tripathi_2011,Testa_2011ApJ, Winebarger_2011, Warren_2012ApJ, giulio_2015A&A}, variability of footpoint emissions~\citep{Testa_2013ApJ...770L...1T,Testa_2014Sci...346B.315T}, time-lag analysis~\citep{Viall_2012ApJ,Viall_2017ApJ...842..108V}, relative amount of spectral line intensities sensitive to high and low temperatures~\citep{Athiray_2019ApJ}. Using the hard X-ray data from the second flight of the Focusing Optics X-ray Solar
Imager (FOXSI-2:\citealp{Glesener_2016SPIE.9905E..0EG}), \cite{Ishikawa_2017NatAs} detected faint but very hot (10 MK) plasma from non-flaring active region (AR), which suggested LF heating.
On the other-hand, a few studies support the HF heating, where plasma is heated in a quasi-steady manner (e.g.,\citealp{Warren_2020ApJ...896...51W}). \cite{Warren_2020ApJ...896...51W} compared the simulated DEM with the derived  DEM of an AR, and their study suggested high-frequency heating.
\cite{Barnes_2021ApJ...919..132B} studied the spatial variation of nanoflare heating of an AR using the DEM and time-lag analysis derived from the EUV observations of Atmospheric Imaging Assembly (AIA:~\citealp{Lemen_2012SoPh}) onboard the Solar Dynamics Observatory (SDO:~\citealp{Pesnell_2012SoPh}). 
Their study suggested that HF nanoflares predominantly influence the core of the AR, while IF nanoflares heat the periphery of the AR.

At present, it is believed that non-flaring ARs are heated by both low and high-frequency heating. However, a thorough investigation of the spatial and temporal evolution of nanoflare heating within ARs is lacking in the literature. Additionally, it is important to determine what fraction of the total energy is deposited to the corona by both HF and LF events.

Addressing these aspects, we investigate the spatial and temporal evolution of nanoflare heating during the lifecycle of AR12758. 
{The spatial distribution here refers to different regions of the active region, composed of multiple coronal loops.}
\cite{Mondal_2023ApJ...955..146M} studied the time-evolution of the temperature and First Ionization Potential (FIP) bias for this AR using the soft X-ray spectra observed by the Solar X-ray Monitor (XSM: \citealp{xsm_flight_performance,XSM_ground_calibration,xsm_XBP_abundance_2021,xsm_microflares_2021}) onboard Chandrayaan-2. This AR is a good case study for heating evolution, as it emerged on the solar disk and decayed in activity during its evolution.

In this study, we compare the evolution of simulated nanoflare-heated plasma emission with observed emissions in X-ray wavelengths. We assume that the AR loops are heated by random nanoflares, with their amplitude and frequency determined from loop properties derived from the linear-force-free (LFF) extrapolation of the photospheric magnetogram observed by the Helioseismic and Magnetic Imager (HMI: \citealp{scherre_2012SoPh}) onboard SDO. This method of studying nanoflare heating has been demonstrated for X-ray Bright Points (XBP) in earlier works by \cite{mondal_2023} and \cite{Mondal_2024ApJ...967...23M}.

To the best of our knowledge, this study is one of the first to investigate the evolution of nanoflare heating frequency throughout the lifecycle of an AR from its emergence to decay on the solar disk. 
The rest of the paper is structured as follows: Section \ref{sec-observation} outlines the observations of AR12758. Section \ref{sec-simulations} describes the detailed simulation setup. The findings are presented and analyzed in Section \ref{sec-results}, followed by a concise discussion in Section \ref{sec-discussion}.

\section{Observations of AR12758}\label{sec-observation}

\begin{figure}[ht!]
\centering
\includegraphics[width=1\linewidth]{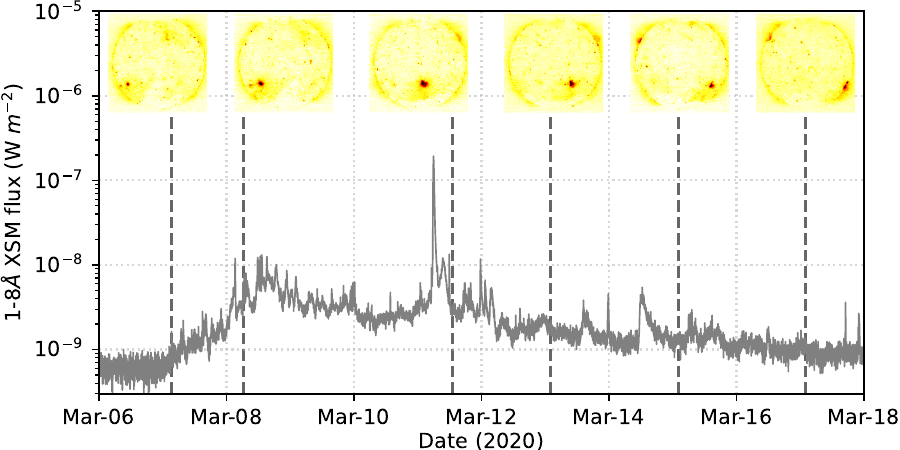}
\caption{1-8~{\AA} X-ray flux variation during the evolution of AR12758. Representative XRT X-ray images are shown in top. Vertical dashed lines indicate the time of the XRT images.}
\label{fig-observation}
\end{figure}

During the minimum of solar cycle 24, AR12758 was formed near the east limb of the Sun on 2020 March 6 and fully emerged after March 8. While crossing the solar-disk, it started decaying and finally went behind the west limb on March 18.
The temporal evolution of AR12758 during this entire period is observed by Chandrayaan-2/XSM\footnote{https://www.prl.res.in/ch2xsm}, which provides the broad-band disk-integrated soft X-ray spectra at every second in 1-15 keV energy range.
The temporal evolution of X-ray flux in 1 to 8~{\AA} is derived from XSM spectra and is shown in Figure~\ref{fig-observation}(grey curve).
Its evolution is also observed by EUV and X-ray imaging instruments SDO/AIA and Hinodel/XRT respectively.
The grey curve in Figure~\ref{fig-observation} shows the 1-8~{\AA} X-ray flux as observed by XSM, during the evolution of this AR. The X-ray images taken by XRT at different times are also shown at the top of Figure~\ref{fig-observation}.

In the absence of any other major activity, the disk-integrated X-rays observed by XSM are dominated by the AR emissions. \cite{Mondal_2023ApJ...955..146M} used the XSM spectroscopic observations to understand the evolution of integrated temperature and FIP bias throughout its evolution on solar-disk. 
In this work, we simulate the evolution of this AR during the non-flaring times, which are compared with the observed X-ray emissions by XSM and XRT.
To simulate the AR emissions we need to know the loop structures, which are obtained from the LFF extrapolation of the observed line-of-sight (LOS) magnetogram by SDO/HMI. 

\section{Simulation setup}\label{sec-simulations}

\begin{figure*}[ht!]
\centering
\includegraphics[width=1\linewidth]{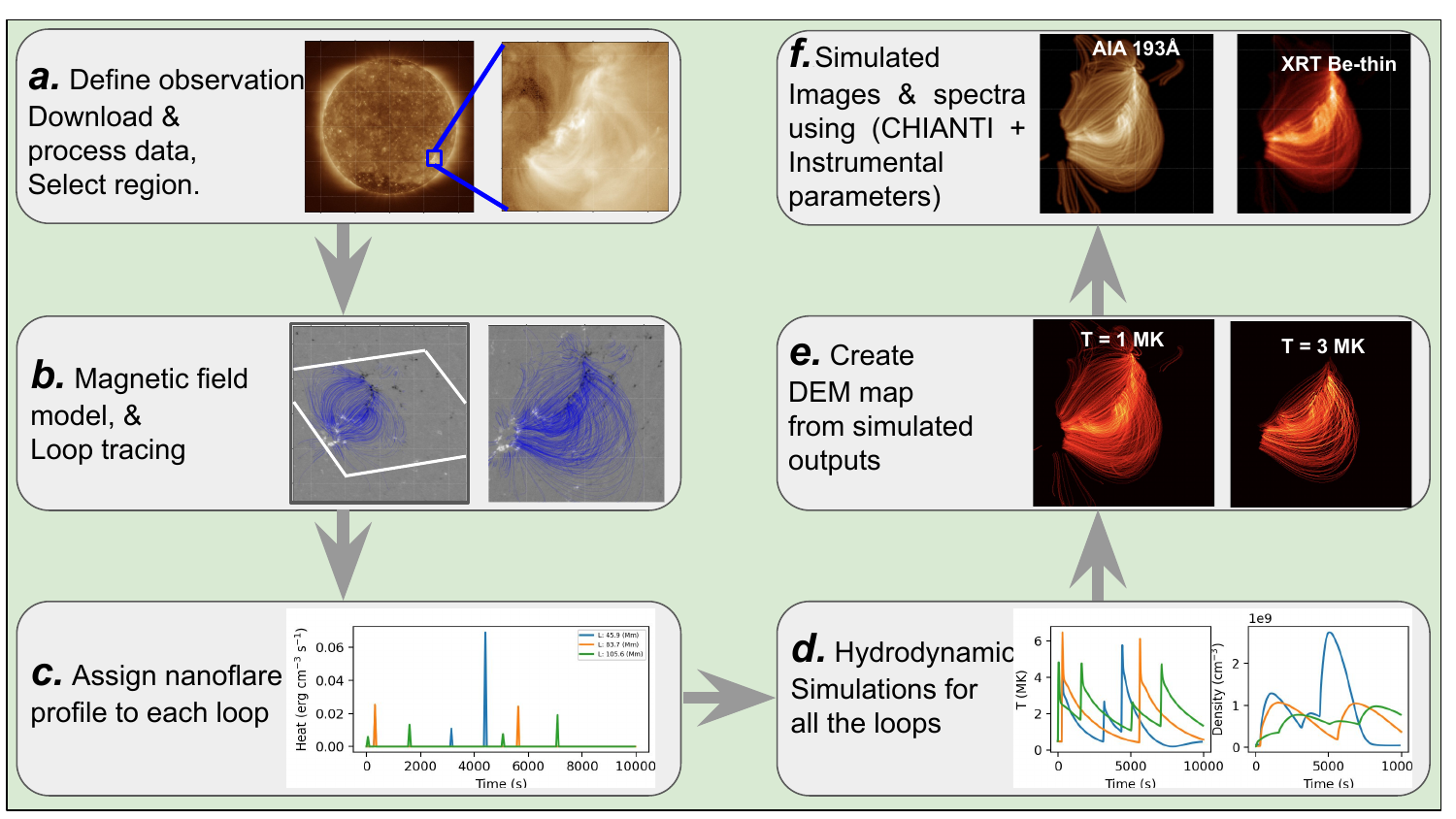}
\caption{Flowchart describing different steps of the simulation setup as discussed in Sections~\ref{step-1}-\ref{sec-sim_image_spec}.}
\label{fig-sunxflowchart}
\end{figure*}

We utilize field-aligned hydrodynamic simulations to synthesize the EUV and X-ray emissions of active region (AR) loops, with the loop structures derived from the extrapolated observed photospheric magnetic fields. Initially, each loop is assumed to be in hydrostatic equilibrium. The hydrodynamic simulations then calculate the plasma response for each loop based on a specified heating profile. To simulate the entire evolution of ARs efficiently, we developed a semi-automated Python pipeline~\citep{mondal_2024_14456559}\footnote{https://github.com/biswajitmb/SunX.git} for easy repetitive setup. Figure~\ref{fig-sunxflowchart} illustrates the step-by-step procedure of the simulation pipeline, with each step detailed in the following sections.

\subsection{Define observation} \label{step-1}

The first step of our pipeline involves downloading the necessary data by specifying a date and time for which solar observations are available from the SDO. The pipeline automatically retrieves the SDO/HMI line-of-sight (LOS) magnetogram and SDO/AIA EUV image for a given passband from the Joint Science Operations Center (JSOC). Once downloaded, the level-1 data is processed for scientific analysis using standard procedures in SunPy~\citep{sunpy_community2020}.
Next, the user is prompted to select a region of interest by using the mouse cursor on the full-disk AIA image, resulting in cutouts of the chosen area. For instance, the blue box in Figure~\ref{fig-sunxflowchart}a indicates the selected region of AR12758 on March 11, 2020, for which we aim to simulate the emission.

\subsection{Magnetic field model}\label{sec-extrapolation}

To model the 3D structure of the magnetic field in the selected region (as described in Section~\ref{step-1}), the pipeline performs a linear-force-free { (LFF:~\citealp{Nakagawa_1972SoPh...25..127N,Seehafer_1978SoPh...58..215S})} extrapolation of the LOS HMI magnetogram. This extrapolation is conducted for multiple values of the force-free parameter ($\alpha$), and the magnetic field lines are traced. The optimal value of $\alpha$ is determined by comparing the traced field lines with AIA images.

For magnetic field line tracing, we utilize the streamline tracing method from the YT package~\citep{YT_package}, specifying random seed locations on the magnetogram where the magnetic field strength exceeds $\pm$20 G. Since the selected region is located away from the disk center, the extrapolated fields may exhibit projection effects in the disk plane. To address this, the observed magnetogram is re-projected with an observer's LOS directed towards the center of the field of view (FOV) using the \verb|reproject_to| functionality of the SunPy \verb|Map| object.

The left panel of Figure~\ref{fig-sunxflowchart}b shows the re-projected magnetogram and the traced field lines from an on-axis observer’s perspective. The white box indicates the selected FOV as observed along the Sun-Earth LOS. To derive the field structure from the Sun-Earth LOS, the traced fields are projected accordingly. The right panel of Figure~\ref{fig-sunxflowchart}b displays the projected field lines on the solar disk, closely matching the observed loop-like structure in the AIA image shown in Figure~\ref{fig-sunxflowchart}a.

Assuming each field line represents a coronal loop, the pipeline estimates the length and length-averaged magnetic field strength ($\langle B \rangle$) for all loops (see Equations 3 and 4 of \citealp{mondal_2023} for details). These parameters are essential for the subsequent step of assigning a nanoflare heating profile for each loop (Section~\ref{sec:heating_function}). {Note that we here use the term ``loop'' to mean the subset of active region magnetic flux that is centered on the selected field line. It does not correspond to an observationally distinct feature in an image. As already mentioned, most of the observed emission from active regions comes from the diffuse component (particularly at low-temperature EUV wavelengths). Distinct loops are modest enhancements over the background.}

\subsection{Nanoflare heating profile}\label{sec:heating_function}

To simulate the coronal loop emissions, we assume the loops are initially in hydrostatic equilibrium with a given length-averaged temperature (here we consider 0.5 MK).
Now we consider that each of the loops are continuously heated by nanoflares~\citep{parker_1988, klimchuk_2015RSPTA}, generated by the release of stored magnetic energy due to the photospheric driving. Here we have assigned a nanoflare heating profile to each loop following the methodology used by \cite{mondal_2023, Mondal_2024ApJ...967...23M}.

The nanoflares heating are represented in terms of a series of symmetric triangular function with time having a half-duration ($\tau$) of 50 s, similar to previous studies, e.g., \citealp{klimchuck_2008ApJ,cargill_2012b,Barnes_2016a}.
The peak heating rate of the nanoflares are randomly selected from the maximum ($H_0^{max}$) and minimum ($H_0^{min}$) possible free energies associated with a loop.
{Whereas, many earlier studies assumes a power-law distribution of the event frequency with their energies (e.g., \citealp{Bradshaw_2016ApJ...821...63B,Tajfirouze_2016ApJ...816...12T}).}
Following \citep{parker_1988}, if $\theta$ is the misalignment angle or Parker angle of a loop from its initial position due to the photospheric driver velocity, the $H_0^{max}$ associated with a loop would be,
\begin{equation}\label{eq_max_energy}
    H_0^{max} = \frac{1}{\tau}\frac{(tan(\theta)\langle B \rangle)^2}{8\pi} (erg cm^{-3} s^{-1}) ,
\end{equation}
where $\langle B \rangle$ is the average coronal field strength along the loop. The minimum possible energy is considered as low as one percent of $H_0^{max}$.

Following \cite{mondal_2023}, the delay time between $l^{th}$ and $(l-1)^{th}$ events associated with a loop is given by:
\begin{equation}\label{eq_delay}
    d = \frac{\tau L}{F} \times H^{l-1}
\end{equation}
Here $F$ is the Poynting flux associated with a loop (half-length is $L$) due to the photospheric driver. This delay-energy relationship is motivated by a physical picture in which magnetic stresses build to a critical level - a critical strand misalignment angle $\theta$ - at which point a nanoflare occurs and releases some of the available free magnetic energy. The more energy that is released, i.e., the stronger the nanoflare, the more time it takes for stresses to again reach the critical level from photospheric driving. 
{This mechanism is similar to an early model proposed by \cite{Rosner_1978ApJ...222.1104R}.}

\subsection{Simulated DEM maps}\label{sec:dem_map}

Once a nanoflare profile is assigned to each loop based on their length and magnetic field strength (Section~\ref{sec:heating_function}), the pipeline executes hydrodynamic simulations for all loops using parallel computing within the available cores of the computer. 
{To solve the hydrodynamic equations, in this study we employ the EBTEL++~\citep{klimchuck_2008ApJ,cargill_2012a,cargill_2012b,Barnes_2016a} 
codes.}
%
EBTEL++\citep{barnes_2024_12675386} is a zero-dimensional model that estimates the time evolution of length-averaged density and temperature for a loop in response to a time-dependent heating profile. In addition to the coronal-averaged density and temperature, EBTEL++ computes the full DEM(T) distribution in the transition region at each time step. We run each simulation run for 10,000 seconds, storing the DEM in the temperature range of 
logT = 5.6 to logT = 7.0 with $\delta$(logT) = 0.1 at a cadence of 10 seconds. We separately average the coronal and transition region DEM(T) over the final 5,000 s. Any influence of the initial conditions disappeared by this time. Each time-averaged DEM(T) represents a snapshot of a bundle of spatially unresolved, randomly heated loop strands.

{Using the coordinates of each loop on the HMI magnetogram, the time-averaged DEM associated with each loop is projected onto a 2D array corresponding to the dimensions of the HMI magnetogram. If multiple loops overlap within a pixel, the average emission from all the loops associated with that pixel is estimated.}
This projection generates a DEM map of the region as a function of temperature. Figure~\ref{fig-sunxflowchart}e displays the DEM map for the selected region at 1 MK and 3 MK, respectively.

\subsection{Simulated images and spectra}\label{sec-sim_image_spec}

Folding the DEM map (Section~\ref{sec:dem_map}) with the temperature response functions of various instrument passbands generates synthetic images that match the plate scale of the DEM map. These synthetic images are then re-binned to align with the actual plate scale of the instrument passbands and convolved with the instrument's point spread function (PSF).

For instance, Figure~\ref{fig-sunxflowchart}f presents synthetic images in the AIA 193 {\AA} and XRT Be-thin passbands. The AIA and XRT temperature responses ($R_i$) are generated using the CHIANTI atomic database \citep{Dere_1997A&AS..chianti,chiantiV10_Zanna2020} with coronal abundances~\citep{Feldman_1992b} via standard routines available in the SolarSoftWare (SSW) package~\citep{Freeland_1998}.
{As the DEM maps are in HMI plate scale, the temperature response functions are converted to the HMI plate-scale of 0.5\arcsec and then the synthetic AIA and XRT images are rebined with their plate-scales of 0.6{\arcsec} and 1{\arcsec} respectively. 
The AIA and XRT images are further convolved with Gaussian PSFs of FWHM 1.2\arcsec\ and 2\arcsec, respectively, and photon noise is applied to them.}

For spectroscopic instruments, the pipeline uses the simulated DEM maps to generate a simulated spectrum based on the given instrument response function. In this study, we simulate the soft X-ray spectrum as observed by the XSM in a disk-integrated (no spatial resolution) observation. An integrated DEM is created from the simulated DEM map. Using this integrated DEM in a multi-component isothermal model and convolving it with the XSM on-axis response function yields the synthetic spectrum. For this purpose, we employ the \verb|chspec| model, a local model within the X-ray spectral fitting package \verb|xspec| \citep{ref-xspec}. The \verb|chspec| model is CHIANTI-based and is described in detail by \cite{biswajit_2021}.

\section{Simulation runs and results}\label{sec-results}

We performed simulations for various evolutionary stages of AR12758, following the methodology outlined in Section~\ref{sec-simulations}. At each time, observed HMI magnetogram is used to assign a random nanoflares heating profile for the AR loops. Following Equations~\ref{eq_max_energy} and \ref{eq_delay}, we also need to provide the Parker angle ($\theta$) and Poynting flux ($F$) associated with the loops. 
The Poynting flux is given by: 
\begin{equation}
F = \frac{1}{4\pi} \left[ v_z B_h^2 - (\mathbf{v}_h \cdot \mathbf{B}_h) B_z \right],
\end{equation}
where $\mathbf{B}$ is the magnetic field and $\mathbf{v}$ is the velocity field.  The subscripts $h$ and $z$ denote the horizontal (or parallel) and vertical (or perpendicular) to photosphere. Using the Differential Affine Velocity Estimator for Vector Magnetograms \citep[DAVE4VM,][]{schuck} to determine $\mathbf{B}$ and $\mathbf{v}$, {we calculate an average Poynting flux in the order of $10^6$ erg cm$^{-2}$ s$^{-1}$ when the active region is near disk center.} For our models, we assume that this constant value applies uniformly across the active region and on all days. {This Poynting flux is lower than the canonical value of $10^7$ erg cm$^{-2}$ s$^{-1}$ given by \cite{Withbroe_1977}, which is consistent with the fact that the active region studied here is magnetically less active, being observed during the minimum of Solar Cycle 24.}

Following \cite{parker_1983}, for an observed magnetic field of 300 G for this AR and a horizontal photospheric driver velocity of 1 km/s, our assumed Poynting flux implies  $\theta = 7^o$. As a first-step, we set the maximum heating rate $H_0^{max}$ and delay $d$ using this constant value of $\theta$ for all days. Later we will allow $\theta$ to vary from day to day, as discussed below.  
{Based on the observed flux of photospheric elemental flux tubes and assuming a typical coronal field strength, \cite{klimchuk_2015RSPTA} estimate the strand radius to be on the order of 100 km. In that case, individual events in our model exhibit average energies on the order of 10$^{24}$ ergs, aligning with the typical energy range of nanoflares (e.g.,~\citealp{parker_1988,Klimchuk_2006SoPh}).}

\begin{figure*}[ht!]
\centering
\includegraphics[width=1\linewidth]{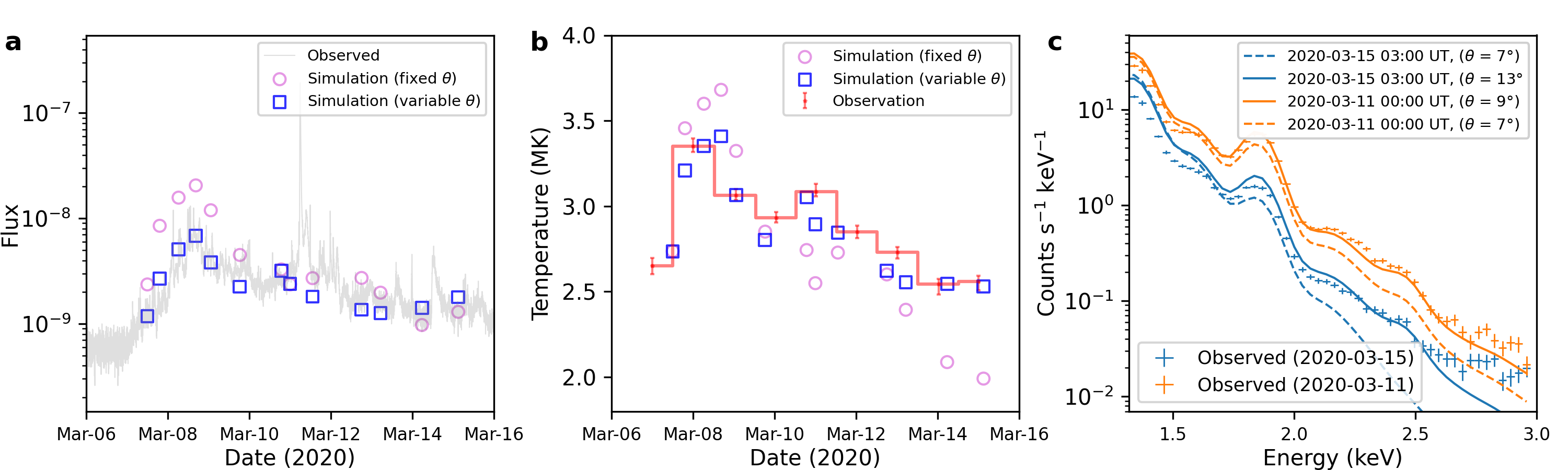}
\caption{(a) Variation of the XSM 1-8 {\AA} X-ray flux (background grey curve) with the simulated flux variation for fixed values of $\theta$ (=7$^o$) (pink-circle) and variable values of $\theta$ (blue square) with time. (b) Variation of observed (red) and simulated (pink-circle and blue-square) temperature derived from X-ray spectra. (c) Observed (error-bars) and simulated (dashed and solid lines) XSM spectra for two representative days of observations. Dashed lines represent the synthetic spectra obtained from the simulation for $\theta$=7$^o$. The solid lines are the synthetic spectra where the $\theta$ values are adjusted for a closer match with the observed spectra.}
\label{fig-compare_xsm}
\end{figure*}

Using the simulated DEM, we synthesise the XSM spectra  (Section~\ref{sec-sim_image_spec}) and estimate the X-ray flux.
The circle points in Figure~\ref{fig-compare_xsm}a represent the simulated evolution of total X-ray flux, compared with the observed flux (grey curve). The absolute values of the simulated flux vary based on the number of loops included. {Each `loop' represents a subset of the active region magnetic flux.  More specifically, each loop represents a collection of magnetic strands that have approximately the same length and are randomly heated by nanoflares with approximately the same energy distribution. The time average of a long duration simulation of a single strand is equivalent to a snapshot of many similar strands that are out of phase.  For the models here we use 2500 loops and normalize the simulated flux by a constant factor of 10$^3$. Normalization is necessary because the simulated strand has an arbitrary cross-sectional area.} A substantially smaller number of loops ($<$500) does not adequately fill the spatial emission of the AR. Interestingly, despite these simplistic assumptions for $F$ and $\theta$, our simulation successfully reproduces the overall variation of the observed X-ray flux. 

Alongside the X-ray flux, we compared the simulated (circle points) and observed (red error bars) temperatures (Figure~\ref{fig-compare_xsm}b) derived from the spectral fitting of the synthetic XSM spectra, {using a method similar to that applied to the observed XSM spectra by \citealp{Mondal_2023ApJ...955..146M}}.
{The synthetic XSM spectra were fitted with an isothermal model, treating temperature and emission measure as free parameters.}
The comparison reveals that the simulated temperatures are systematically overestimated during the active region's emerging phase and underestimated during its decay phase.
To investigate the reasons for these deviations, we compared the observed and simulated spectra. Figure~\ref{fig-compare_xsm}c displays the comparison of observed (error bars) and simulated (dashed lines) spectra for two representative days. The analysis reveals that the shapes of the simulated spectra, which determine the temperature, differ significantly from the observed spectra.

{In the next step, we varied the spatially averaged value of $\theta$ at different evolutionary stages of the AR in the simulation, which influences the spectral shape, to ensure alignment between the simulated spectra and the observations over time.}
For instance, the solid curve in Figure~\ref{fig-compare_xsm}c shows the simulated spectra after adjusting $\theta$ from its initial value of 7° to better match the observed data. Consequently, the simulated flux variations, represented by the blue square points in Figure~\ref{fig-compare_xsm}a, exhibit a closer alignment with the observed flux. Throughout the evolution of the AR, we found that $\theta$ varies between 6° and 14°. From this point forward, we will present results based on the varying $\theta$ model. Note that the variable $\theta$ applies to Equation~\ref{eq_max_energy}. We maintain a constant Poynting flux.

\begin{figure*}[ht!]
\centering
\includegraphics[width=1\linewidth]{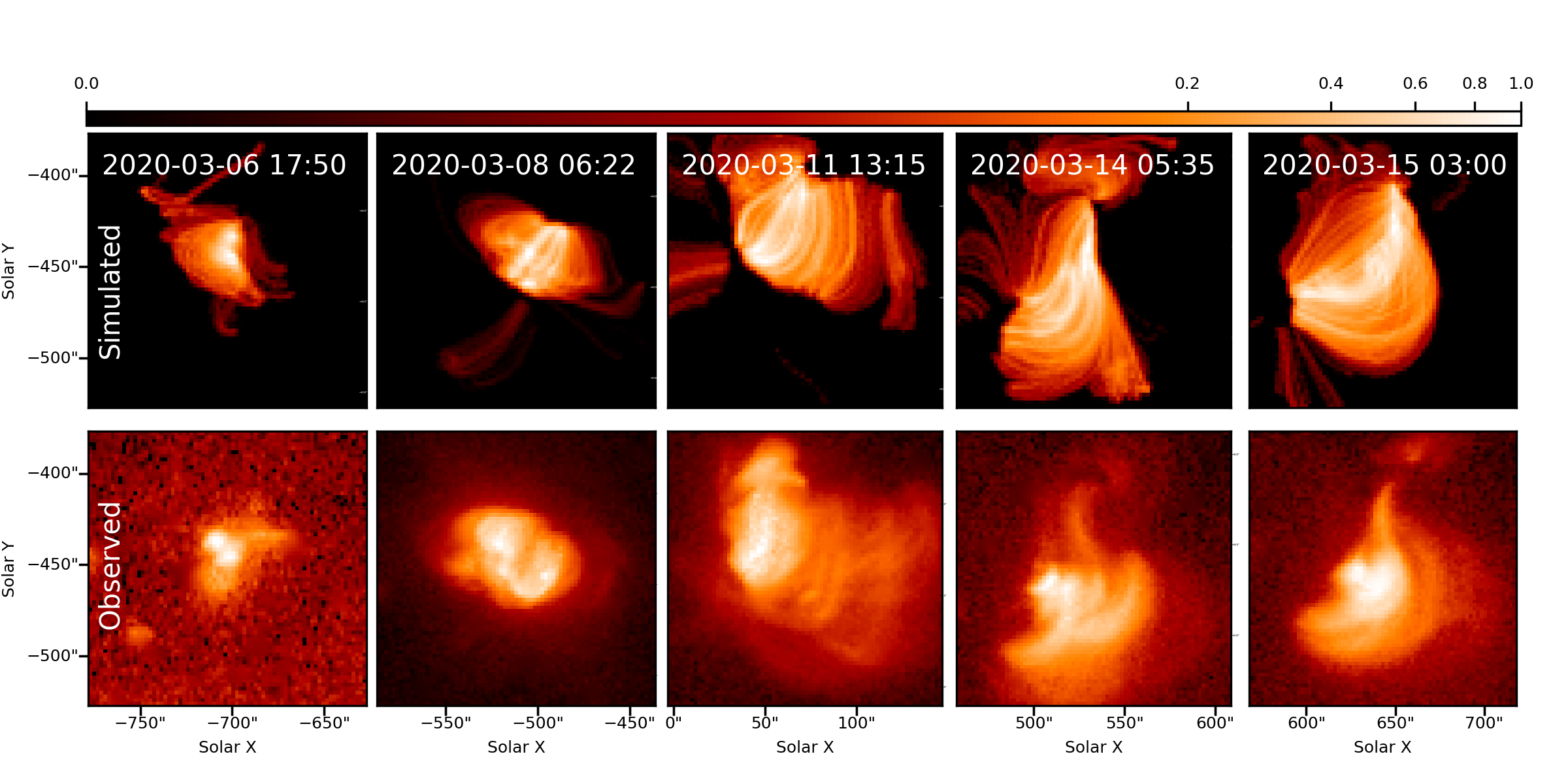}
\caption{Simulated (top row) and observed (bottom row) X-ray images in XRT Be-thin filter during the evolution of AR12758.}
\label{fig-xrt}
\end{figure*}

Figure~\ref{fig-xrt} illustrates the spatial evolution of simulated and observed X-ray emissions using the XRT Be-thin filter. Given the simplicity of the model, a one-to-one match between the spatial variations of brightening in the simulated and observed images is not expected. 
{Nevertheless, the overall simulated emission geometry throughout the evolution of the AR are not far away from the observations.
The simulated images include only the resolved coronal loops, whereas the observed images also capture emissions from the unresolved diffuse corona.
}

\begin{figure*}[ht!]
\centering
\includegraphics[width=1\linewidth]{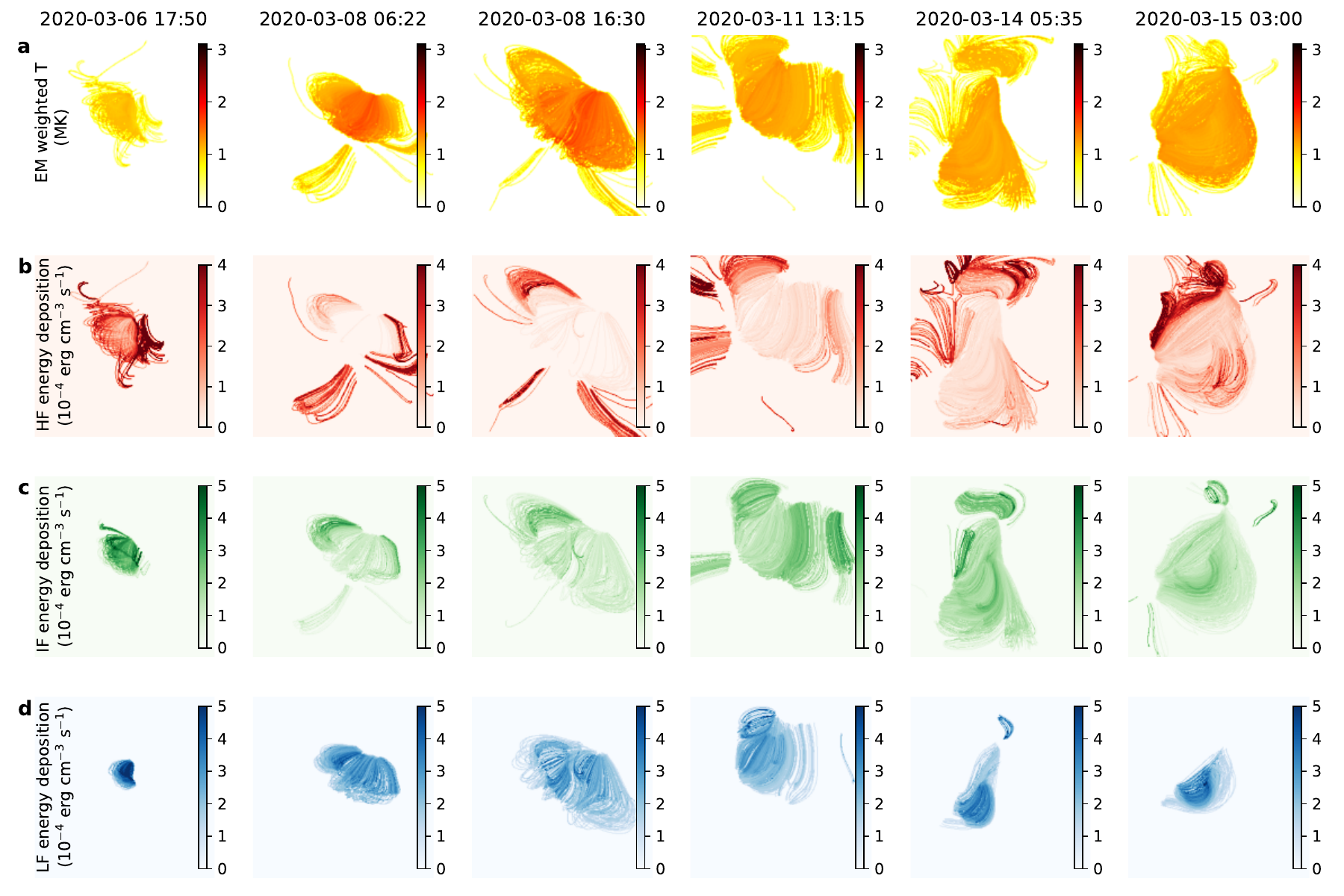}
\caption{Panel-a show the spatial distribution of EM-weighted temperature throughout the evolution of the AR. Panel b-d, show the distribution of the average energy deposited by HF, IF, and LF nanoflares.}
\label{fig-heatdeposite}
\end{figure*}

In this study, we focus on examining the spatial and temporal variations of heating parameters. In the first step, we analyze the variation of simulated EM-weighted temperature along the line of sight, as shown in the top row of Figure~\ref{fig-heatdeposite}. 
During the emerging phase of the AR, the center exhibits higher temperatures compared to the peripheral regions. {Conversely, during the decay phase, the temperature decreases to become more uniformly across the entire AR. }This suggests significant differences in the properties of the heating between these phases.

To understand the heating contributions from different nanoflares, we analyzed the spatial variation of heat deposited by HF, IF, and LF nanoflares. The classification of these events is based on the following limits:
\begin{align*}
    HF: t_{rep} < 0.5t_{cool}\notag\\
    IF: 0.5t_{cool} < t_{rep} < 2t_{cool}\notag\\
    LF: t_{rep} > 2t_{cool}
\end{align*}
Here, $t_{rep}$ is the repetition time between two nanoflares and $t_{cool}$ is the cooling time of the plasma heated by a nanoflare. 
Following \cite{2015ApJ...799..128L} the above criterion can be expressed in terms of the temperature change of the cooling plasma in response to a nanoflare event. The temperature evolution after a nanoflare can be approximately described by an exponential function: $T(t) = T_0exp(-t/t_{cool})$, where $T_0$ is the maximum temperature during the event, and $t_{cool}$ is the cooling-time; when $T(t_{cool}) = T_0/e$. Thus,
\begin{align*}
    HF: {T(t_{rep})} > 0.61{T_0} \\
    IF: 0.14{T_0} < {T(t_{rep})} \leq 0.61{T_0} \\ 
    LF: {T(t_{rep})} \leq 0.14{T_0}
\end{align*}
Based on these criteria, we classify the heating events for each loop and create maps of HF, IF, and LF energy deposition using their coordinates. For overlapping loops at the same coordinate, we calculate the average energy deposition for each heating category. Figures~\ref{fig-heatdeposite}b-d display the heat deposition maps for HF, IF, and LF events. While all types of nanoflares contribute to heat deposition, their spatial contributions change throughout the evolution of the active region, as discussed in Section~\ref{sec-freq}.

\section{Discussion and Conclusion}\label{sec-discussion}

We aim to investigate the spatial and temporal evolution of nanoflare heating frequency during the lifetime of AR12758, from its emergence to decay on the solar disk. Using a field-aligned hydrodynamic model that incorporates nanoflare heating scenarios, we simulate the emission of the active region. Nanoflare frequency and magnitude are determined by loop properties derived from observed SDO/HMI magnetograms. The simulated emissions are then compared with observed data. In this section, we discuss the primary outcomes of our study.

\subsection{Temporal variation of X-ray flux}

Figure~\ref{fig-compare_xsm}a compares the temporal variation of the integrated X-ray flux of the AR, observed by XSM (grey curve), with the simulation results (circles and squares). The circle points use the corresponding SDO/HMI magnetograms at each time, with all other simulation parameters kept constant.
Despite the simplistic assumptions, the simulation successfully reproduces the observed flux variation over time, highlighting the crucial role of the magnetogram in determining heating properties. 
{These results are consistent with findings reported by \cite{Ugarte-Urra_2017ApJ...846..165U, Ugarte-Urra_2019ApJ...877..129U}, which demonstrate that AR EUV emissions are correlated with their magnetic activity. Their study also shows that steady heating can partially reproduce the magnitude of EUV radiance.
However, we found that matching the spectrally integrated radiance with observations does not necessarily imply that the heating properties can explain emissions at different energy ranges. For instance,} the synthesized XSM spectra (dashed curves in Figure~\ref{fig-compare_xsm}c) fail to reproduce the observed spectral shape, resulting in significant deviations in the simulated temperatures compared to observations, as shown in Figure~\ref{fig-compare_xsm}b.

In the next simulation setup, we adjusted the average critical angle ($\theta$) of the loops over time to match the spectral shape and temperature. The synthetic spectra, shown by the solid curves in Figure~\ref{fig-compare_xsm}c, now align more closely with the observed flux (Figure~\ref{fig-compare_xsm}a) and temperature (Figure~\ref{fig-compare_xsm}b). We found that $\theta$ ranges from 6° to 14° throughout the AR's evolution. These values are consistent with the estimates of 14° by \cite{parker_1988} and 10° by \cite{klimchuk_2015RSPTA}, considering coronal energy losses for a typical AR.

A caveat in our model is the assumption of a Poynting flux ($F$) that is constant in time and uniform across the AR. In reality, $F$ may vary both temporally and spatially. A thorough investigation is needed to explore how these variations impact the spatial distribution of emissions. Additionally, we assume a constant average value of $\theta$ for all loops at any given time. However, in reality, $\theta$ may vary at different spatial locations within the AR. 
Our current understanding of what determines $\theta$ is limited \citep{klimchuk_2015RSPTA}, though a promising new idea is current sheet loss of equilibrium~\citep{Klimchuk_2023FrP}.

\subsection{Spatial variation of emissions and temperature}
 
Simulated XRT Be-thin emissions exhibit greater brightening in the core of the AR, while the periphery shows less brightening, consistent with the observed emission geometry (Figure \ref{fig-xrt}). The XRT Be-thin emission geometry is highly dependent on the nanoflare heating frequency, as described in Figure 8 of \cite{Mondal_2024ApJ...967...23M}. Thus, a closer match with observations indicates that the heating frequency is adequate.

Figure \ref{fig-heatdeposite}a shows the time evolution of emission measure (EM) weighted temperature maps. Since the cool ($\sim$1 MK) EM dominates, the weighted temperature is more sensitive to cooler emissions, though this does not imply the AR lacks very hot plasma~($>$5 MK). Early in the emerging phase (e.g.,~March~06), the AR is dominated by cooler plasma ($\sim$1 MK). However, by March 08, the core becomes hotter than the periphery. As the AR ages, temperature distribution becomes more uniform. 
Most of the hot emissions are confined to the core of the AR, consistent with observations (e.g., \citealp{zanna_2022ApJ...934..159D, rao_2023ApJ...958..190R}). Figure 5 of \cite{zanna_2022ApJ...934..159D} shows the temperature structure of AR12759 (similar activity as AR12758) derived from XRT Be-thin/Al-poly filter ratios, where the hot emission concentrated towards the center.

In our model, most of the higher temperatures in the AR core are associated with low-frequency (LF) heating (Section~\ref{sec-freq}) for two main reasons.
First, at low frequencies, loops have sufficient time to drain before the next heating event, allowing the same energy input to heat the lower-density plasma to higher temperatures than would be possible in denser plasma. Second, LF heating predominates in the core of the active region, where the coronal magnetic field is stronger, and nanoflares are more energetic.

\subsection{Heating frequency}\label{sec-freq}

Throughout the evolution of the AR, heat is deposited
by all types of nanoflares, 
{(Figure~\ref{fig-heatdeposite}) which agrees with earlier studies (e.g., \citealp{Bradshaw_2016ApJ...821...63B}).}
Figure 5 shows that the core of the active region spends most of its time in a LF heating state, the periphery spends most of its time in a HF heating state, and the region in between spends most of its time in an IF heating state. This suggests a correlation with loop length. We can understand the dependence as follows. 

{The heating frequency ($\nu$) depends on the repetition time between successive events, $d$. It also depends on the cooling time, but for simplicity we ignore this dependence. The multitude of small magnetic strands that comprise the coronal magnetic field are constantly twisted and tangled by complex photospheric motions. Nanoflares are initiated at the electric current sheets separating the strands whenever a critical level of stress is reached. One plausible picture is that this critical stress corresponds to a critical misalignment angle between the strands.

The repetition time is the time required for photospheric driving to replenish the magnetic energy released by the nanoflare, i.e., $d = \frac{\Delta E}{dE/dt}$,
where $\Delta E \propto (\langle B \rangle \tan \theta)^2/8\pi$ is the maximum nanoflare energy density, and $dE/dt = F / L$ is the energy density buildup rate with Poynting flux $F$ and loop halflength $L$. As a general rule, the magnetic field strength decreases with loop length:  $\langle B \rangle \propto L^{-\delta}$. \cite{Mandrini_2000ApJ} find that $\delta = 0.9$ is a reasonable representation of active regions. Putting this together, we get $\nu = 1/d \propto L^{0.8}$ for the heating frequency. Thus, frequency increases with loop length, as clearly seen in Figure~\ref{fig-heatdeposite}. If the current sheet loss of equilibrium idea is correct to determine the value of critical angle ($\theta$), then $\tan \theta \propto 1/L$ ~\citep{Klimchuk_2023FrP} and the frequency dependence on loop length is even stronger:  $\nu \propto L^{2.8}$. We note that our loops have constant cross section, and a full analysis would require that a possible length dependence of the loop expansion factor be taken into account.}


Earlier studies suggested that active region cores experience HF heating based on the observed steady nature of the EUV/X-ray emissions (e.g., \citealp{Warren_2011ApJ}). This, however, ignores the possibility of spatially unresolved dynamic behavior which is the basis of the nanoflare picture. Our study indicates that the AR core heating is dominated by LF/IF nanoflares depending on the AR age.
\cite{Barnes_2019}, and \cite{Barnes_2021ApJ...919..132B} predicted nanoflare heating frequency for an AR using average 12-hour EUV observations from SDO/AIA combined with a different modeling approach, suggesting that HF events predominantly heat most of the AR, particularly the core. This contradicts our findings, which may stem from differences in the activity levels of the ARs studied or from observational biases, as most past studies relied on EUV observations, which are more sensitive to cooler plasma. Differences could also be due to the differences in the heating models. \cite{Barnes_2019} assume a power law energy distribution, $N(E) \propto E^{-2.5}$, whereas we assume a flat random distribution. Thus, they have a much higher proportion of weak events, and the events occur much more frequently to achieve the same Poynting flux. Our maximum event energy is given by the free magnetic energy associated with the small-scale twisting and tangling, which, for $\theta = 7^o$, is 1.5\% of the total magnetic energy. \cite{Barnes_2019} assume a maximum event energy equal to the potential magnetic energy, which is much larger. Finally, our delay time between two events is proportional to the energy of the first event. This corresponds to a magnetic energy ceiling (critical stress/angle). Their delay time is proportional to the energy of the second event. This is appropriate if there is a magnetic energy floor, i.e., all events relax the field to a ground state.
{A similar heating model is used by \cite{Bradshaw_2016ApJ...821...63B}, whose study suggests a broad heating frequency, with the cooling time in the AR core being small compared to the longer loops in the AR periphery.}
To reach definitive conclusions, more detailed studies of the evolution of multiple ARs with varying activity levels are necessary.

\begin{figure*}[ht!]
\centering
\includegraphics[width=1\linewidth]{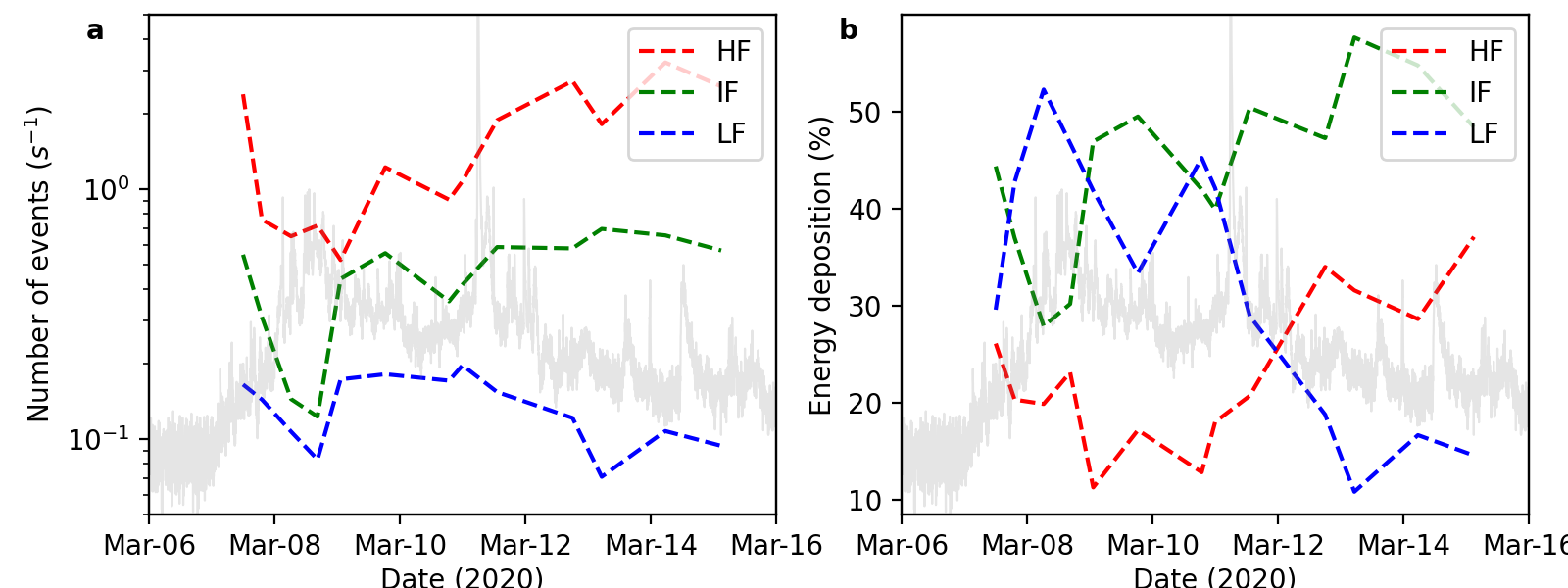}
\caption{Number of events (a) and fraction of the total energy (b) deposited by high-frequency (HF, red), intermediate-frequency (IF, green), and low-frequency (LF, blue) events. The background gray curve shows the normalized X-ray flux observed by XSM throughout the evolution of AR 12758.}
\label{fig-hatfract}
\end{figure*}

To understand the contribution of different type of nanoflares in global AR heating budget,
Figure~\ref{fig-hatfract}a shows the spatially averaged, time evolution of HF, IF, and LF events over the AR's lifetime. While HF events dominate in terms of occurrence rate, their contribution to overall coronal heating is relatively less. 
To quantify this, Figure~\ref{fig-hatfract}b shows the fraction of energy deposited by different events. During the emerging phase, LF events dominate energy deposition. After full emergence (e.g., March 9-11), energy is deposited almost equally by both IF and LF nanoflares. As the AR ages (e.g., after March 12), the energy contribution from LF events decreases, while the contributions from IF and HF events increase.

Determining heating frequency from observations can be biased by the wavelength used; for instance, EUV observations are more sensitive to cool emissions, while X-ray observations favor hot emissions. Accurately determining heating frequency requires observations that can detect both high and low temperatures. Current imaging and spectroscopic instruments have limitations in detecting very hot plasma \citep{Winebarger_2012}. To address this, efforts have been made to constrain hot plasma, such as through the Marshall Grazing Incidence X-ray Spectrometer (MaGIXS:~\citealp{Champey_2022JAI}) rocket flights, the upcoming CubeSat Imaging X-ray Solar Spectrometer (CubIXSS:~\citealp{Caspi_2023SPD....5420704C}), {and upcoming Multi-slit Solar Explorer (MUSE:~\citealp{Pontieu_2022ApJ...926...52D}).}
The first successful flight of MaGIXS demonstrated its capability to detect hot plasma \citep{Athiray2024A} and determine heating frequency \citep{Mondal_2024ApJ...967...23M}, although it did not observe any hot ARs. The very recent second successful flight of MaGIXS is expected to provide a better understanding of AR heating frequency. However, continuous measurements throughout the entire evolution of ARs are needed to constrain the heating model throughout the AR's lifecycle.

\section{Summary}\label{sec-summary}

We have studied the temporal and spatial evolution of nanoflare heating frequency in the lifecycle of AR12758 by combining field-aligned hydrodynamic simulations and observations. Our findings indicate that all types of nanoflares (HF, IF, and LF) are present, with HF nanoflares dominating the number of events per unit time. However, HF nanoflares primarily contribute to cool plasma and account for only a small fraction of the total energy deposition. In the AR's emerging phase, energy deposition is dominated by LF nanoflares, while both LF and IF nanoflares share energy deposition post-emergence. As the AR ages, IF and HF nanoflares become more dominant. 
The spatial distribution of heating frequencies shows a distinct pattern: the core of the active region spends most of its time in a LF heating state, the periphery is dominated by HF heating, and the area between the core and periphery experiences IF heating.
To determine whether this scenario is typical for AR heating, further studies involving multiple ARs with varying activity levels are essential. Although IF and LF events are significant contributors to hot plasma ($>$3 MK), their emissions are much lower compared to the cooler plasma produced by HF events. Therefore, understanding nanoflare contributions to coronal heating should account for both observational biases and the energy deposition by these events, along with their frequency.

\acknowledgments{BM's research was supported by an appointment to the NASA Postdoctoral Program at the NASA Marshall Space Flight Center, administered by Oak Ridge Associated Universities under contract with NASA. The work of JAK was supported by the GSFC Heliophysics Internal Scientist Funding Model (competitive work package) program and by the NASA Living With a Star program. }
{We acknowledge the helpful comments from an anonymous reviewer to improve the manuscript.}

{\textit{Facilities:} SDO(AIA, HMI), Hinode (XRT), Chandrayaan 2 (XSM).}

{\textit{Software:} Astropy~\citep{Astropy2018AJ....156..123A}, IPython~\citep{ipython_2007CSE.....9c..21P},matplotlib~\citep{matplotlib_2007CSE.....9...90H}, NumPy~\citep{2020NumPy-Array}, scipy~\citep{2020SciPy-NMeth},
SunPy~\citep{sunpy_community2020}, SolarSoftware~\citep{Freeland_1998}, EBTEL++~\citep{Barnes_2016a}, Chianti~\citep{Dere_1997A&AS..chianti,chiantiV10_Zanna2020}}


\newpage
\bibliography{myref}   
\bibliographystyle{aasjournal}

\end{document}